\documentstyle[preprint,aps,epsfig]{revtex}
\tightenlines  
\begin{document} 
\title{\LARGE \bf One-parameter Darboux-transformed quantum actions in 
Thermodynamics}

\author{Haret C. Rosu\footnote{E-mail: rosu@ifug3.ugto.mx;
Fax: 005247187611.}
  and Fermin Aceves de la Cruz}
\address{Instituto de F\'{\i}sica IFUG, Apdo Postal E-143, Le\'on, Gto, Mexico}


\maketitle
\thispagestyle{empty}

\begin{abstract}
{\scriptsize
We use nonrelativistic supersymmetry, mainly Darboux
transformations of the general (one-parameter) type,
for the quantum oscillator thermodynamic actions. Interesting Darboux
generalizations of the fundamental
Planck and pure vacuum cases are discussed in some detail with relevant plots. 
It is shown that the one-parameter Darboux-transformed Thermodynamics refers
to superpositions of boson and fermion excitations of positive and negative 
absolute temperature, respectively.
Recent results of Arnaud, Chusseau, and Philippe
regarding a single mode 
oscillator Carnot cycle are 
extended in the same Darboux perspective. We also conjecture
a Darboux generalization of the fluctuation-dissipation theorem.}
\end{abstract}

\vspace{5mm}

\section{Introduction} 

Recently, Arnaud, Chusseau, and Philippe (hereafter ACP) \cite{acp} 
studied the work done per Carnot cycle by a single mode oscillator of the 
ideal $LC$ type operating between
two baths of different temperatures. They used 
the old (1906) prescription of Einstein of discrete $\hbar \omega$ exchanges 
with the baths. Within this approach, they confirmed the Carnot theory of the 
efficiency of cyclic engines for this case. As known, the energy of the
oscillator is given by Planck's distribution
\begin{equation} \label{1}
U_P(\omega, \beta)=\frac{\hbar \omega}{2}+\frac{\hbar \omega}
{\exp (\beta \hbar \omega)-1}~.
\end{equation}
One can define an action function
\begin{equation} \label{2}
f_{P}(x)=\frac{U_P}{\omega}=\frac{\hbar}{2}+\frac{\hbar}
{\exp (\hbar x)-1}=\frac{\hbar}{2}{\rm coth}(\frac{\hbar x}{2})~,
\end{equation}
where $x=\beta \omega$.
This action plays the role of a generalized force in the process of
frequency change.
In order to evaluate the efficiency, ACP introduced
a `two-variable entropy'
\begin{equation} \label{3}
s(x,y)\equiv xf_{P}(y)-\int ^{x}f_{P}(z)dz=\frac{\hbar x}{2}
{\rm coth}(\frac{\hbar y}{2})-\ln \sinh (\frac{\hbar x}{2}) +C~,
\end{equation}
with the property $s(x,x)=s(x)$, the latter being the usual entropy. 
Employing standard thermodynamic formalism,
ACP calculated the efficiency of both reversible and nonreversible Carnot
cycles.

ACP made the interesting remark that the oscillator action
$f_{P}(x)$ fulfills as particular solution the following Riccati equation
\begin{equation} \label{4}
\frac{df}{dx}+f^2=\left(\frac{\hbar}{2}\right)^2~.
\end{equation}
As a matter of fact, the constant vacuum action $f_V=\frac{\hbar}{2}$ is also
a solution of the same Riccati equation, whereas the
pure thermal action
$f_T=\frac{\hbar}{\exp (\hbar x)-1}$ is a particular solution of 
the following equation
\begin{equation} \label{5}
\frac{df}{dx}+\hbar f +f^2=0~,
\end{equation}
which is a Bernoulli equation (or a particular type of Riccati 
equation).
Here we would like to 
study the thermodynamic consequences of using methods belonging to supersymmetric 
quantum mechanics \cite{review} focusing on  one-parameter 
Darboux transformations of thermodynamic actions.

\bigskip

\section{ General solution of the supersymmetric partner Riccati equation} 

\bigskip

In Witten's supersymmetric quantum mechanics \cite{W}, which is a simple 
application of Darboux transformations \cite{review}, one usually starts with a 
known particular solution of
a Riccati equation without linear term
\begin{equation} \label{6}
\frac{df_{p}}{dx}+f^2_{p}=V_1(x),
\end{equation}
that we call the bosonic Riccati equation. We are not interested here in 
a so-called factorization 
constant that can be placed at the right hand side. The function $V_1$ is an 
exactly solvable potential for the Schr\"odinger equation at zero energy
\begin{equation}\label{7}
\left(\frac{d}{dx}+f_p\right)\left(\frac{d}{dx}-f_p\right)w_b=
\frac{d^2w_b}{dx^2}-V_1(x)w_b=0~.
\end{equation}
The particular solution $w_b$ is usually called a bosonic zero mode. 
It is connected to the Riccati
solution through $f_p=\frac{1}{w_b}dw_b/dx$.
Next, changing the 
sign of the first derivative in Eq.~(\ref{6}) one calculates the outcome 
$V_2(x)$ using the same Riccati solution
\begin{equation} \label{6a}
-\frac{df_{p}}{dx}+f^2_{p}=V_2(x)~.
\end{equation}
We call the latter equation the fermionic Riccati equation.
The function $V_2(x)$ is known as the supersymmetric 
partner of the initial potential $V_1(x)$. The corresponding zero mode fulfills
\begin{equation}\label{7a}
\left(\frac{d}{dx}-f_p\right)\left(\frac{d}{dx}+f_p\right)w_f=
\frac{d^2w_f}{dx^2}-V_2(x)w_f=0~.
\end{equation}

On the other hand, already in 1984, Mielnik \cite{M} studied
the ambiguity of the factorization of the Schr\"odinger equation for the 
oscillator that led him to the general (one-parameter-dependent) 
solution of the companion Riccati equation for that case. In other words,
one looks for the general solution \cite{review} 
\begin{equation} \label{8}
  - \frac{d f_{g}}{dx}+ f_{g}^2 =-\frac{d f_p}{dx}+f^2_p=V_2(x)~.
\end{equation}
The latter equation can be solved for $f_{g}$ by employing the Bernoulli ansatz
$ f_{g}(x)= f_p(x) - \frac{1}{v(x)}$, where $v(x)$ is 
an unknown function \cite{Z}. One obtains for the function $ v(x)$ the
following Bernoulli equation
\begin{equation} \label{9}
 \frac{dv(x)}{dx} + 2v(x)\, f_p(x) =1~,
\end{equation}
that has the solution
\begin{equation} \label{10}
 v(x)= \frac{{\cal I}_{0b}(x)+ \lambda}{w_b^{2}(x)}~,
\end{equation}
where $ {\cal I}_{0b}(x)= \int _{0}^{x} \,
w_b^2(y)\, dy$,
and we consider $\lambda$ as a positive integration constant that is 
employed as a free parameter.

Thus, the general fermionic Riccati solution is a one-parameter function
$ f_{g}(x; \lambda)$ of the following form
\begin{equation} \label{13}
 f_{g}(x;\lambda)=  f_p(x) -
 \frac{d}{dx}
\Big[ \ln({\cal I}_{0b}(x) + \lambda) \Big]=  \frac{d}{dx}
\Big[ \ln \left(\frac{w_b(x)}{[{\cal I}_{0b}(x) +
\lambda]}\right)\Big]=\frac{d}{dx}\ln w_b(x;\lambda)~
\end{equation}
where $w_b(x;\lambda)=\frac{w_b(x)}{{\cal I}_{0b}(x) +\lambda}$.
The range of the $\lambda$ parameter is conditioned by ${\cal I}_{0}(x) +
\lambda \neq 0$ in order to avoid singularities. This is a well-known 
restriction \cite{M}.
According to the supersymmetric construction, one can use
this general fermionic Riccati solution to calculate a one-parameter family
of bosonic potentials as follows
\begin{equation} \label{gen}
\frac{df_{g}}{dx}+f_g^2=V_{1,g}~,
\end{equation}
where 
\begin{equation} \label{pot}
V_{1,g}=V_1-2\frac{d^2}{dx^2}\ln\left({\cal I}_{0b}(x) +\lambda\right)
\end{equation}
enters the linear equation
\begin{equation}\label{Sgen}
\left(\frac{d}{dx}+f_g\right)\left(\frac{d}{dx}-f_g\right)w_b(x;\lambda)=
\frac{d^2w_b(x;\lambda)}{dx^2}-V_{1,g}w_b(x;\lambda)=0~.
\end{equation}

In the limit $\lambda 
\rightarrow \infty$, Eq.~(\ref{gen}) goes into Eq.~({\ref{6}) 
because $f_g\rightarrow f_p$ and $V_{1,g}\rightarrow V_{1}$. 

One can think of Eq.~(\ref{gen}) as a 
generalization of the thermodynamic Riccati equation (\ref{4}). Of 
interest are the one-parameter oscillator actions $f_g$ rather than the 
`potentials' $V_{1,g}$. Since in supersymmetric quantum mechanics $V_{1,g}$
are the general Darboux-transformed potentials, we shall call the $f_g$ as
Darboux-transformed actions.   

\bigskip

\subsection{The Planck case}
Using $f=w'/w$ in Eq.~(\ref{4}), 
where the 
prime denotes the derivative with respect to $x$, leads to the linear equation
\begin{equation}
w^{''}-\left(\frac{\hbar}{2}\right )^2w=0~,
\end{equation}
having the particular zero-mode solution $w_a=W_a\sinh(\frac{\hbar}{2}x)$.
Thus, the general Riccati solution is a one-parameter function
$ f_{gP}(x; \lambda)$ of the following form
$$
 f_{gP}(x;\lambda)=  f_P(x) - 
 \frac{d}{dx}
\Big[ \ln({\cal I}_{0a}(x) + \lambda) \Big]
$$
\begin{equation}
=  \frac{d}{dx}
\Big[ \ln \left(\frac{w_{a}(x)}{[{\cal I}_{0a}(x) +
\lambda]}\right)\Big].
\end{equation}
Accordingly, the two-variable entropy will also become a parameter-dependent 
function
\begin{equation}
s_g(x,y;\lambda)\equiv xf_{gP}(y;\lambda)-\int ^{x}f_{gP}(z;\lambda)dz~,
\end{equation}
where from all the basic calculations as performed by ACP can be 
easily repeated. For example, to calculate Carnot efficiencies one can 
use the generalized ACP formula
\begin{equation} \label{Carnot}
\eta _{C,g}=1-\frac{T_{cold}}{T_{hot}}
\frac{s_g(b,a;\lambda)-s_g(u;\lambda)}{s_g(a;\lambda)-s_g(v,u;\lambda)}~,
\end{equation}
and the same values of the parameters as in ACP, 
i.e., $T_{cold}=1/4$, $T_{hot}=1$,
$a\equiv\beta_{hot}\omega _{1}=1$, $b\equiv\beta_{cold}\omega _{2}=4$,
$u\equiv\beta_{cold}\omega _{3}=2c$, and $v\equiv\beta_{hot}\omega _{4}=2$,
keeping $c$ as a free parameter. 
The results of this subsection are illustrated in the plots of Figs.~(1) - (4).

\bigskip

\subsection{The vacuum case}
For this case, the  
particular Riccati solution is the vacuum action $f_p=f_V=\frac{\hbar}{2}$. The 
corresponding zero mode is $w_V\propto e^{\hbar x/2}$. 
The usual entropy $s_V(x,x)$ of the vacuum fluctuations is zero as a result 
of a simple calculation, whereas the 
modified entropy has a kink-like behavior between the $\frac{\hbar}{2}$
(bosonic) solution and the $-\frac{\hbar}{2}$ (fermionic) 
solution. Plots of this case are 
displayed in Figs.~(5) - (8).

\bigskip

\subsection{The symmetric zero mode: Fermi-Dirac action at negative T}
The cases in {\bf A} and {\bf B} could be considered particular cases
of the general zero-mode $w_g=Ae^{\hbar x/2}+Be^{-\hbar x/2}$ that can be also
used as solution in Eq.~(\ref{7}). 
The Planck action corresponds to $A=-B=\frac{1}{2}$ (antisymmetric zero-mode), 
while the vacuum case to $A=$ arbitrary and $B=0$. One can use any other
type of zero-modes. For example, the symmetric zero-mode
$w_s=W_s\cosh (\frac{\hbar}{2} x)$ (see Figs.~(9) - (10)) is an interesting case
since if we trace back to the action we get 
\begin{equation}
f_s=-\frac{\hbar}{2} +\frac{\hbar}{\exp (-\hbar x)+1}~,
\end{equation}
i.e., a Fermi-Dirac action for negative $x$.
We have shown in a previous paper that
the $\lambda$ parameter is equivalent to the quotient $A/B$ \cite{boya}.
Thus, the general Riccati solution introduces effects of the second 
linear independent solution. The problem then turns into a subtle 
interpretation of the mathematical results. 
We have found at least one possible physical significance.
For the Planck case, the second linear independent zero-mode is exactly
the $\cosh$
function and the corresponding action is the aforementioned Fermi-Dirac action
of negative $x$. 
We attach now the minus sign to the temperature parameter
in $x=\omega /T$ and recall that 
the issue of negative absolute (spin) temperatures first appeared in Physics in 
1951 when Purcell and Pound were able to produce sudden reversals of the 
direction of an external magnetic field applied to a crystal of LiF \cite{PP}.
Since then many other experiments with negative temperatures have been
devised in nuclear spin systems and the corresponding  
`violations' of the second law of thermodynamics 
were a subject of discussion \cite{ram}.
Thus, the one-parameter Darboux transformations of the Planck action
are a way of introducing upon it
the effect of a Fermi-Dirac action of negative absolute temperature.
On the other hand, in the figures (9) - (10) 
we introduce effects of the Planck action
of positive temperature on the Fermi-Dirac action of negative temperature.

\bigskip

\section{The Darboux Generalization of the Fluctuation-Dissipation Theorem}
 
Another interesting application refers to dissipative $RLC$ systems 
where the oscillator action enters
Nyquist-Johnson spectral power noise distributions of the type 
(fluctuation-dissipation theorem \cite{N})
\begin{equation}
P(\omega ,\beta)=\frac{\omega}{\pi}R(\omega ,\beta)f_P~.
\end{equation}
One can think of the corresponding generalization
\begin{equation}
P(\omega , \beta ;\lambda)=\frac{\omega}{\pi}R(\omega ,\beta)f_g(\beta \omega;
\lambda)
\end{equation}
and hope to study even in simple experiments the 
significance of the present approach predicting a Darboux generalization of 
the fluctuation-dissipation theorem.

\bigskip

\section{ Conclusion} 

In this work, Planck's thermodynamic oscillator action  
is generalized to a one parameter Darboux
family of actions. We also consider the bosonic vacuum case separately in 
the same way.
The Planck action and the pure vacuum case correspond to the asymptotic 
limit of the Darboux parameter $\lambda \rightarrow \infty$. 
In the Planck case, all the other $\lambda$
cases correspond to a system made of bosons at temperature $T$
interacting with an equal system of fermions at temperature $-T$.
In the vacuum case, the $\lambda \neq \infty$ cases describe the interaction
of the bosonic and fermionic vacua. 
The efficiencies of the ideal oscillator Carnot
cycles based on the Darboux-modified Planck and vacuum entropies
are calculated along the lines described
by Arnaud, Chusseau and Philippe. Systems of negative Kelvin temperatures are
hotter than those of positive $T$ \cite{ram} and therefore they always 
represent the hot bath. In real, dissipative cases, the same type
of generalization is suggested for the fluctuation-dissipation theorem. 
Finally, we mention that a multiple-parameter Darboux generalization is also
possible \cite{mult}.

\nopagebreak

\bigskip
\section*{Acknowledgment}

\noindent

The first author benefited from 
useful electronic discussions with Prof. Jacques Arnaud and Dr. Laurent Chusseau.



\nopagebreak


\bigskip


\newpage


\vskip 2ex
\centerline{
\epsfxsize=180pt
\epsfbox{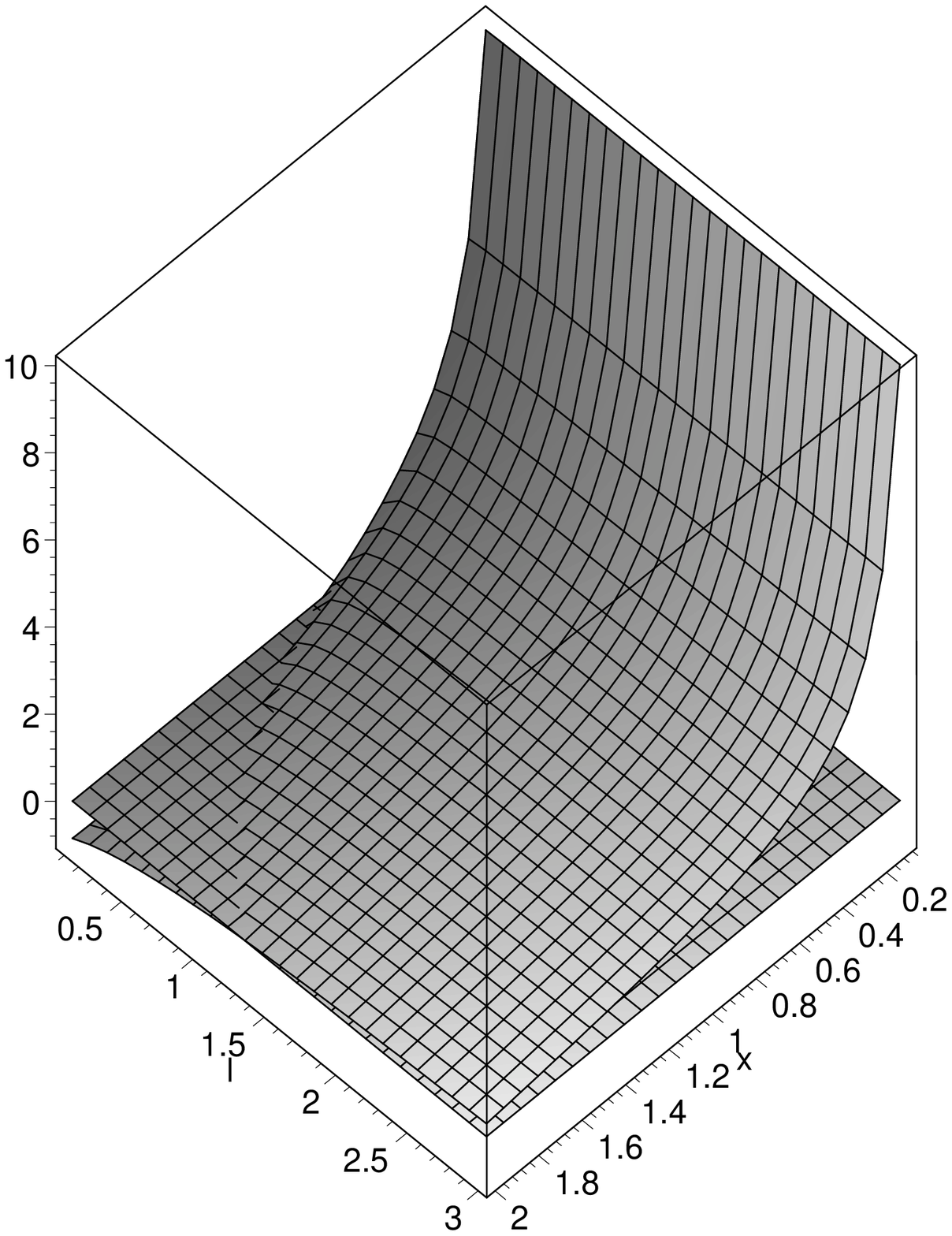}}
\vskip 4ex
\noindent
{\small{Fig. 1}.$\quad$
The one-parameter Darboux-modified Planck action $f_g(x; \lambda)$ as a 
function of $x$ and $\lambda$.}

\vskip 2ex
\centerline{
\epsfxsize=180pt
\epsfbox{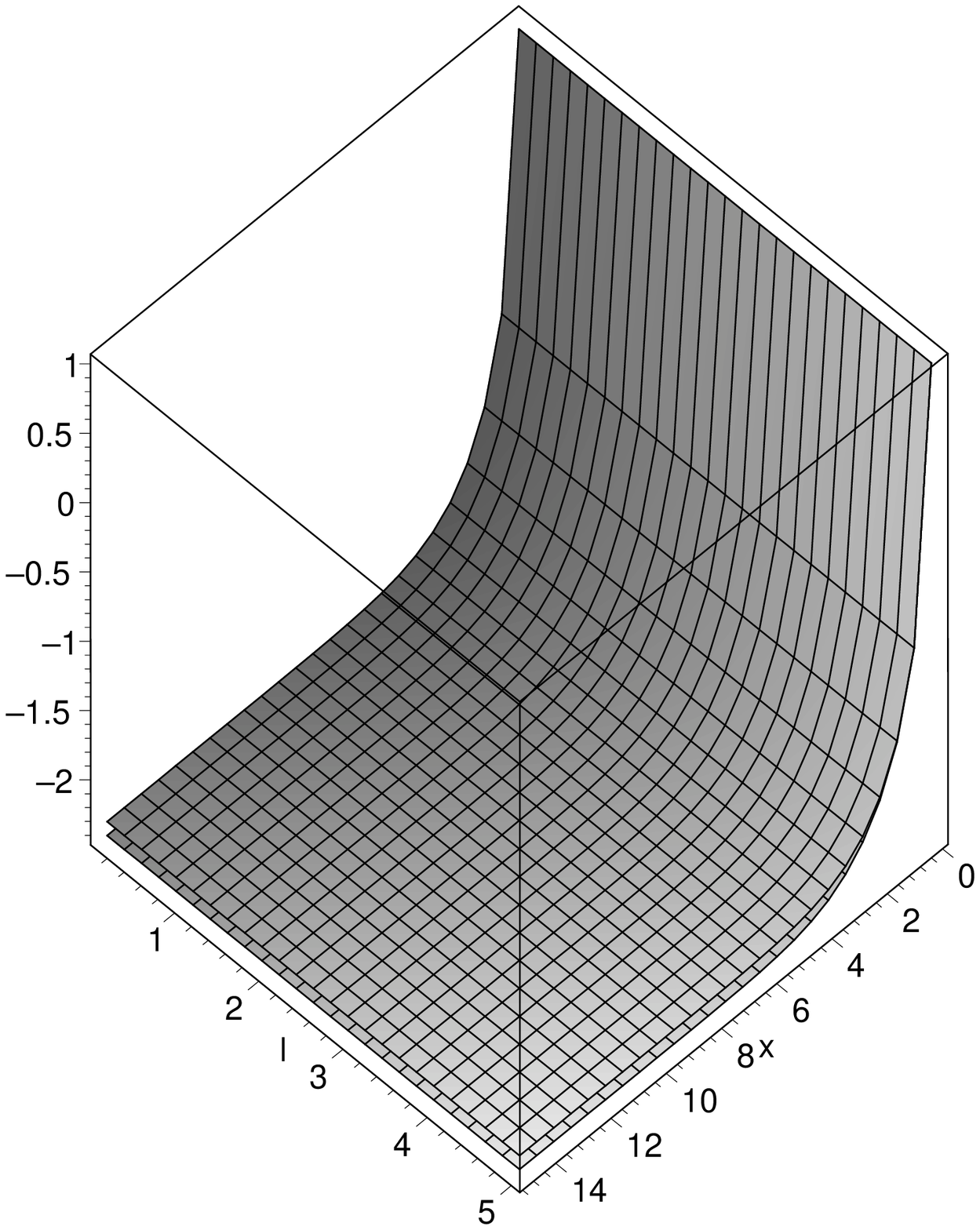}}
\vskip 4ex
\noindent
{\small{Fig. 2}.$\quad$
The standard and the one-parameter entropy functions. There are only small
differences between them. For more details see Fig.~3.}

\vskip 2ex
\centerline{
\epsfxsize=180pt
\epsfbox{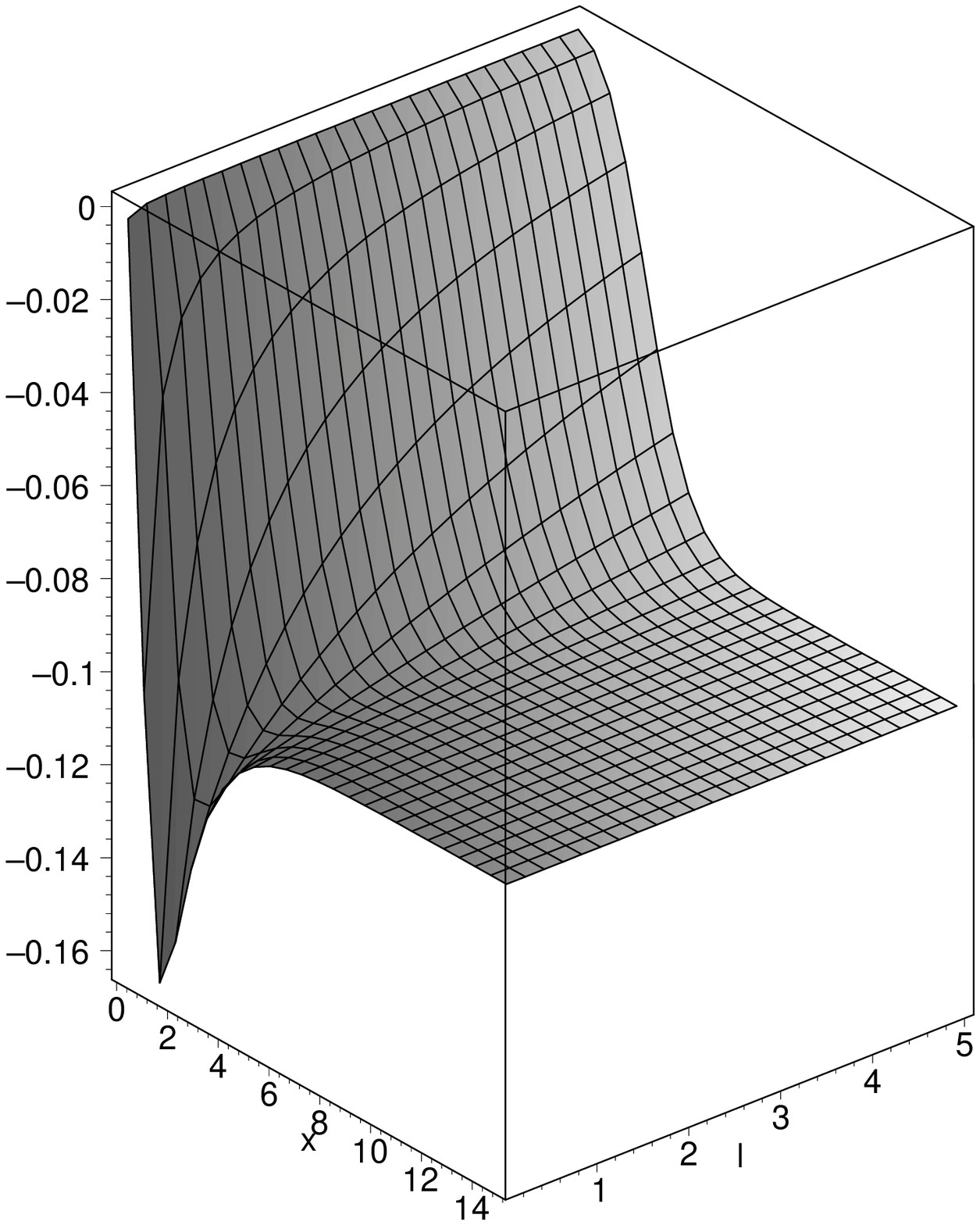}}
\vskip 4ex
\noindent
{\small{Fig. 3}.$\quad$
The difference between entropies $\Delta s=s_g-s_P$.}


\vskip 2ex
\centerline{
\epsfxsize=150pt
\epsfbox{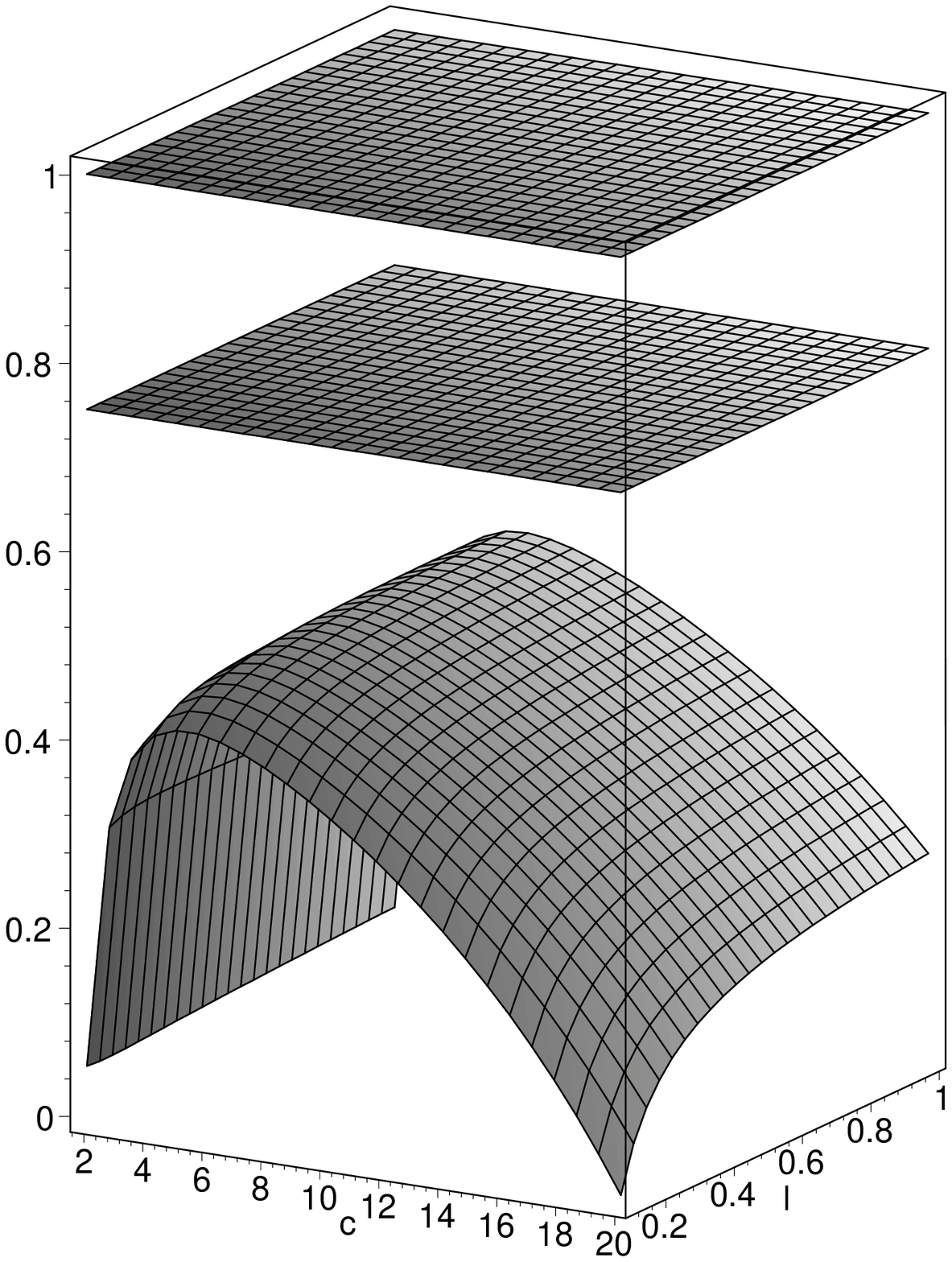}}
\vskip 4ex
\noindent
{\small{Fig. 4}.$\quad$
Darboux-modified Carnot efficiencies for the Planck case. The plane at the 
height 0.75 corresponds to the maximum Carnot efficiency for the parameters 
used by ACP.}


\vskip 2ex
\centerline{
\epsfxsize=180pt
\epsfbox{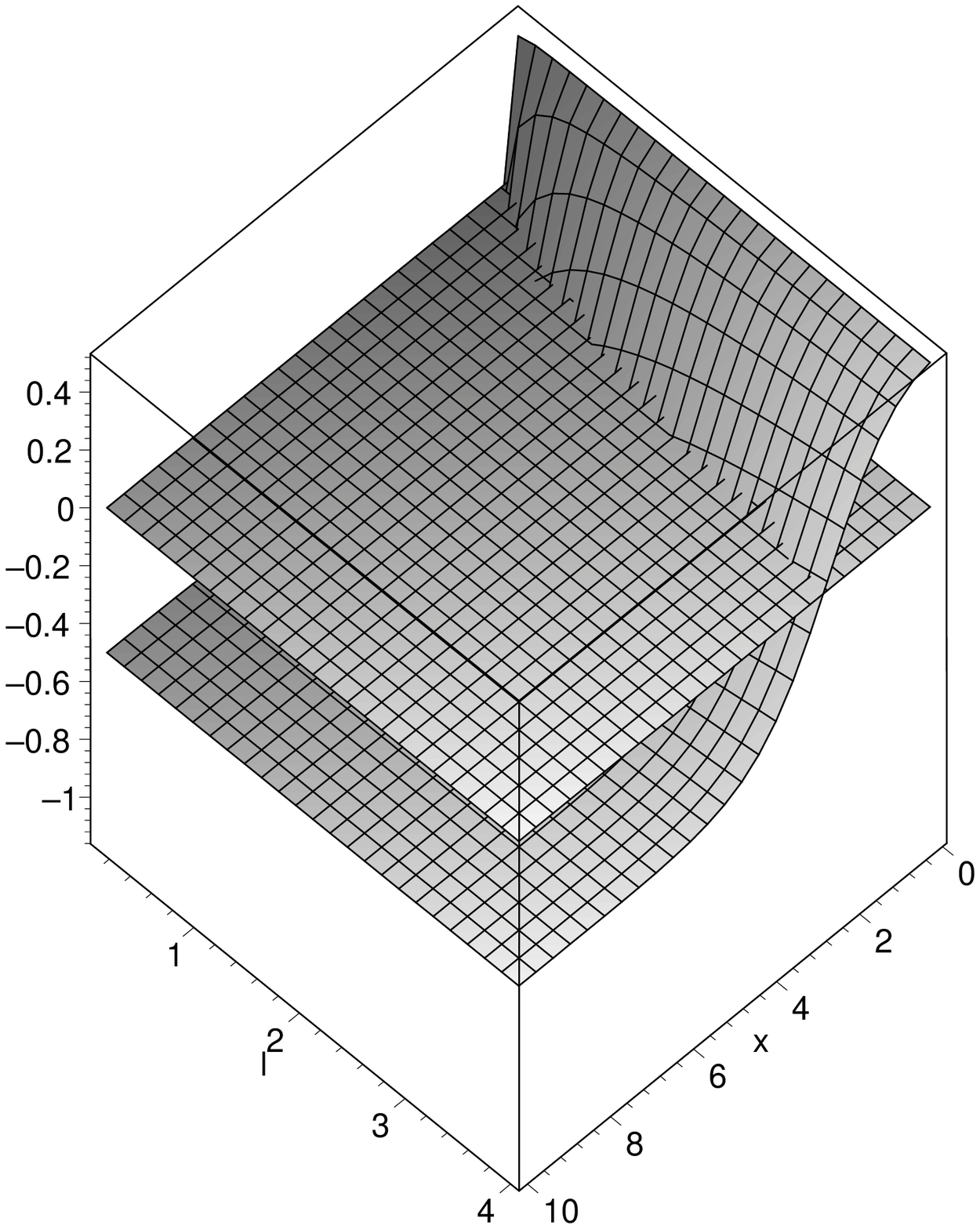}}
\vskip 4ex
\noindent
{\small{Fig. 5}.$\quad$
The Darboux-modified zero-point action. It starts at the normal $\frac{\hbar}{2}$ value
at $x=0$ and goes to $-\frac{\hbar}{2}$ at large values of $x$. The shape is
that of a usual kink (switching) function between $\pm \frac{\hbar}{2}$
for large values of the parameter. For small $\lambda$ values see Fig.~6.}


\vskip 2ex
\centerline{
\epsfxsize=180pt
\epsfbox{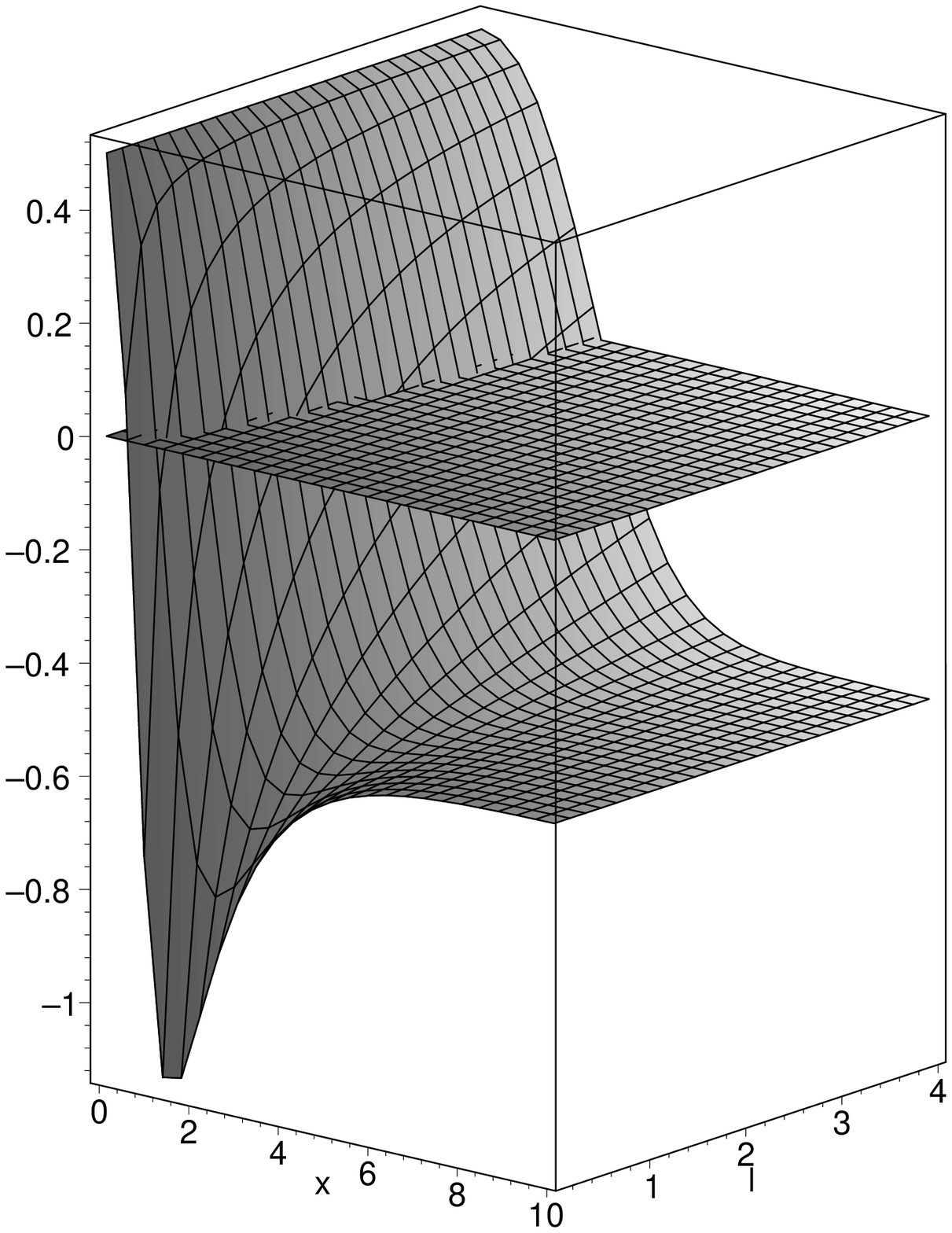}}
\vskip 4ex
\noindent
{\small{Fig. 6}.$\quad$
The same as in the previous figure but for a different orientation to emphasize
that at low values of $\lambda$ the general Riccati thermodynamic vacuum kink 
deviates from the common definition of a kink and even turns singular for 
$\lambda _c=(x-\sinh x)/2$.}

\vskip 2ex
\centerline{
\epsfxsize=180pt
\epsfbox{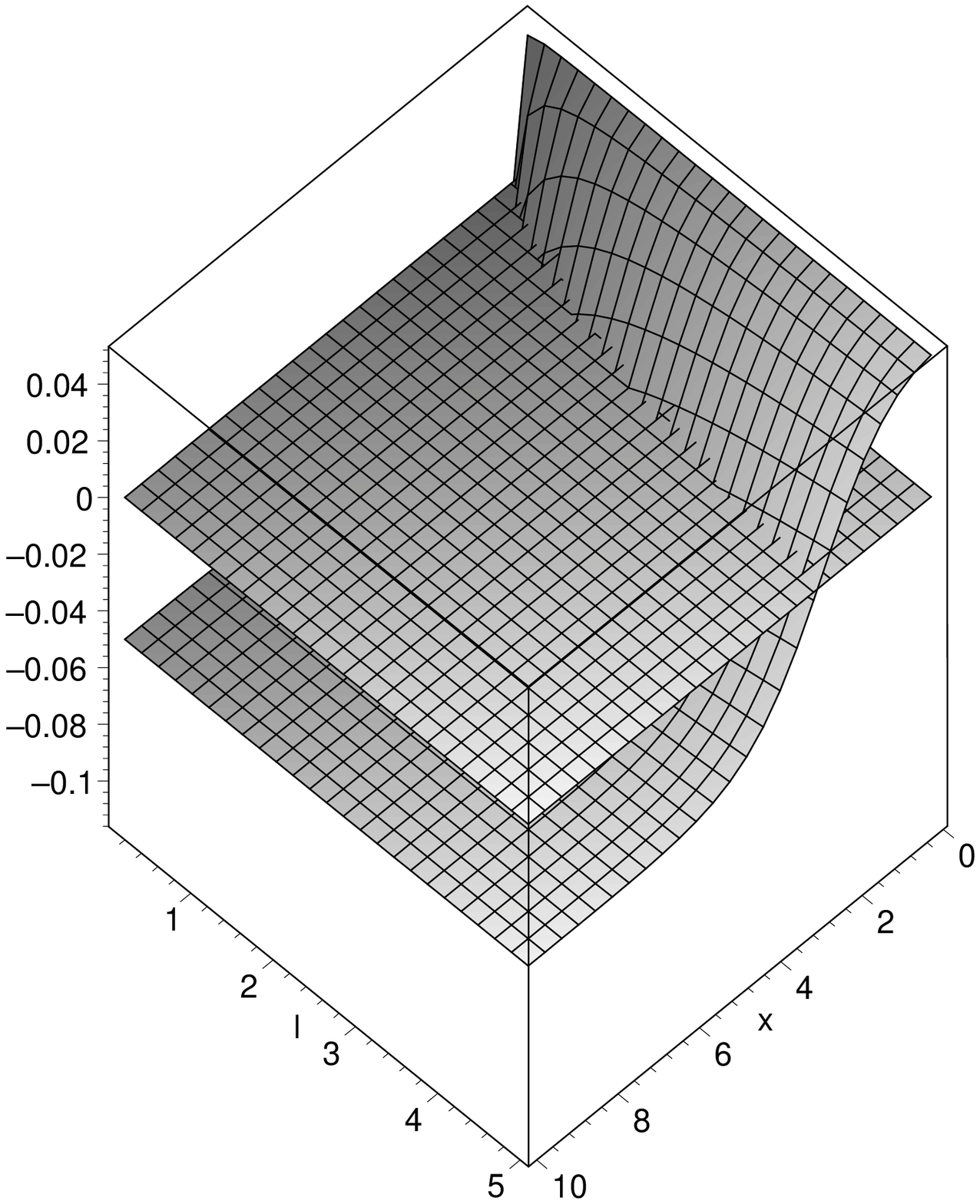}}
\vskip 4ex
\noindent
{\small{Fig. 7}.$\quad$
The usual vacuum entropy and the Darboux-modified one.}

\vskip 2ex
\centerline{
\epsfxsize=180pt
\epsfbox{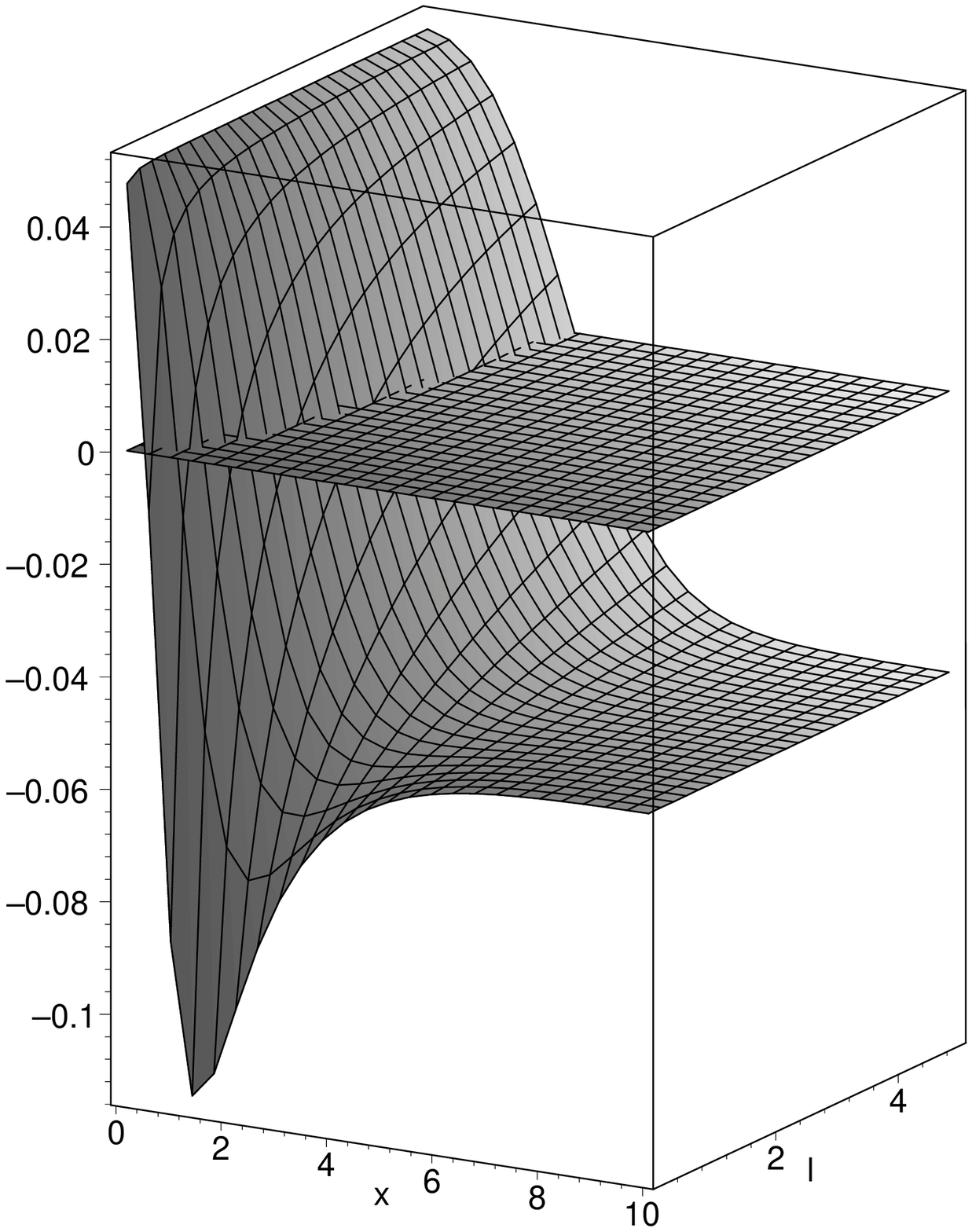}}
\vskip 4ex
\noindent
{\small{Fig. 8}.$\quad$
The same as in the previous figure but for a different orientation.}


\vskip 2ex
\centerline{
\epsfxsize=180pt
\epsfbox{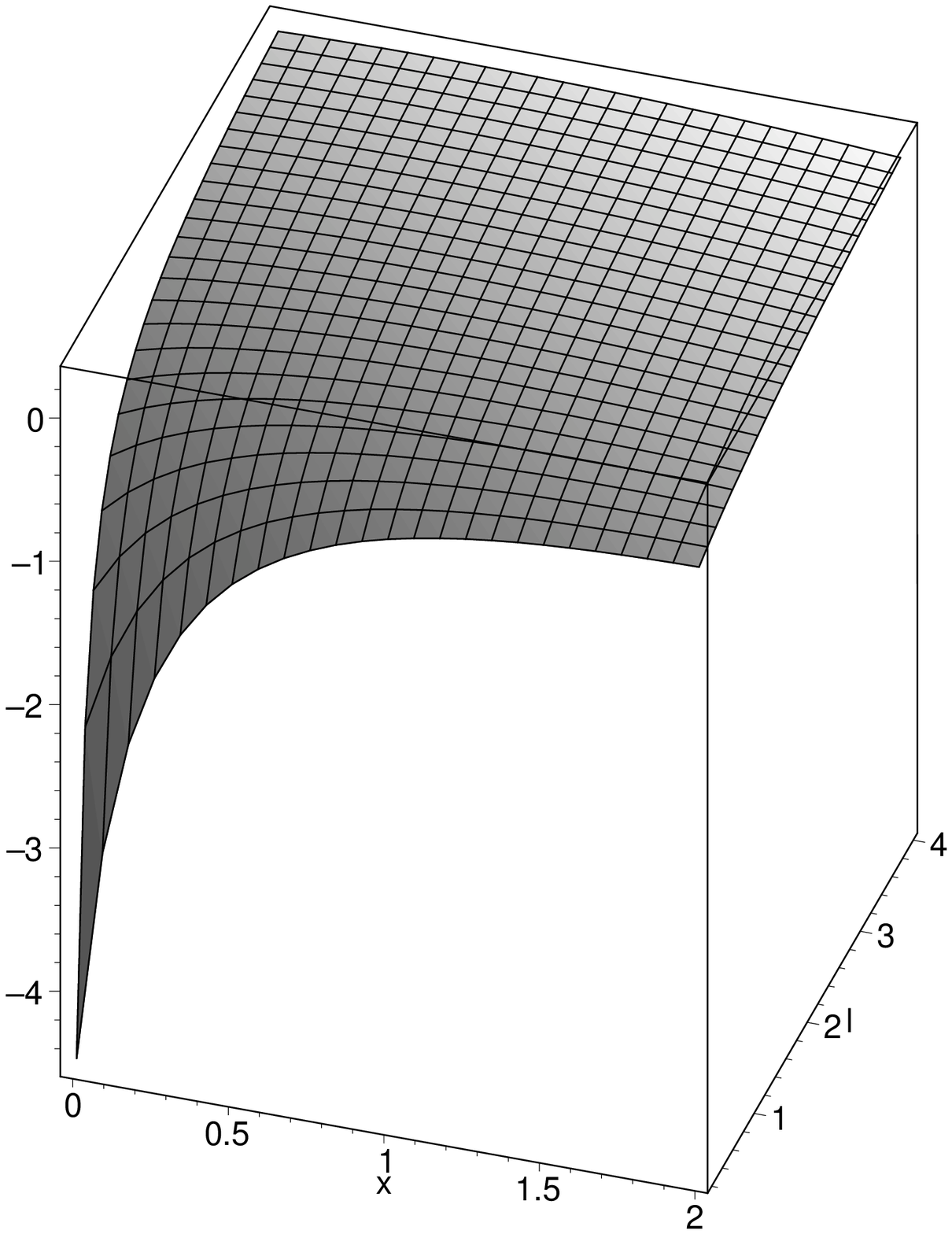}}
\vskip 4ex
\noindent
{\small{Fig. 9}.$\quad$
The one-parameter Darboux-modified Fermi-Dirac action of negative T.}

\vskip 2ex
\centerline{
\epsfxsize=180pt
\epsfbox{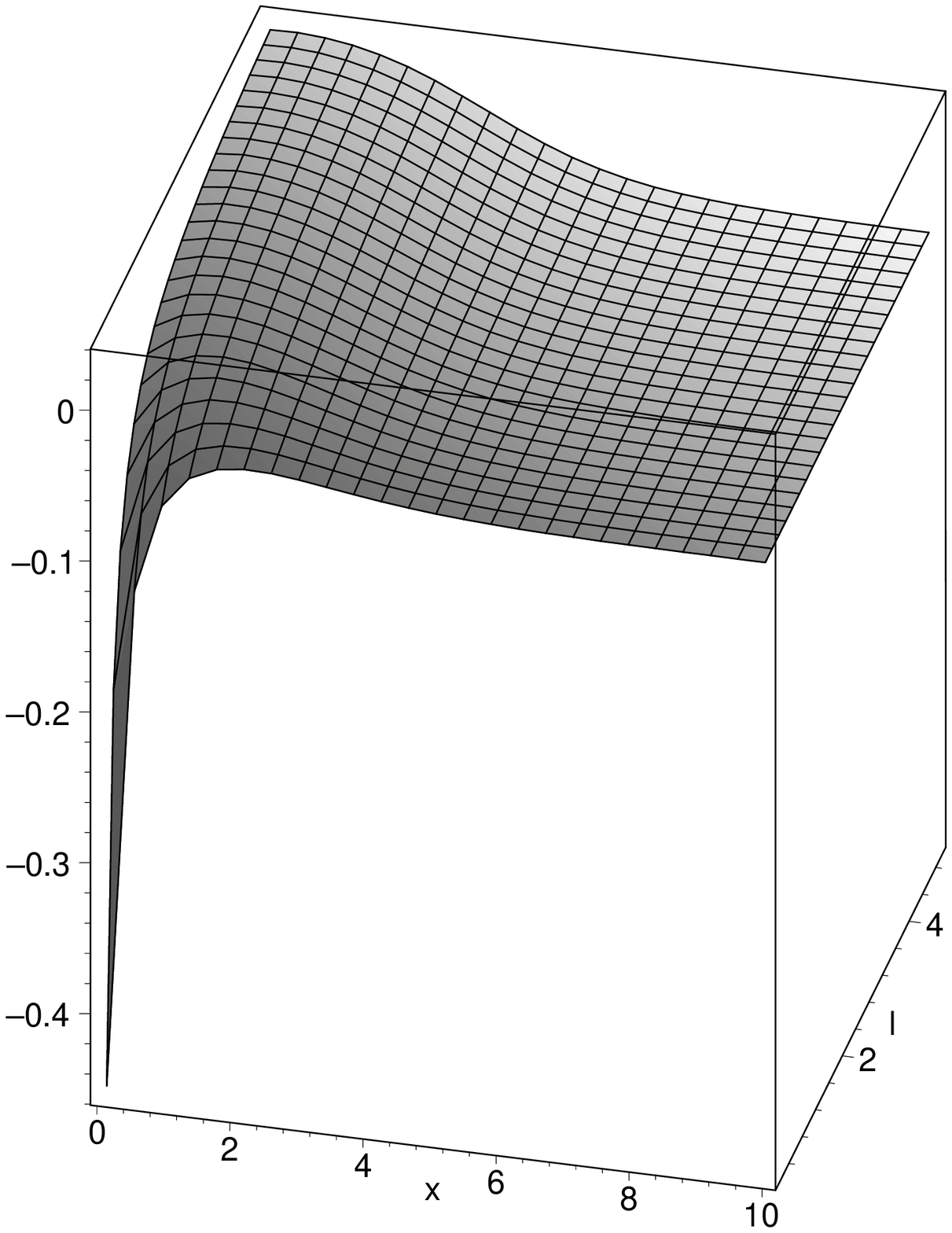}}
\vskip 4ex
\noindent
{\small{Fig. 10}.$\quad$
The one-parameter Darboux-modified Fermi-Dirac entropy of negative T.}

\newpage
\begin{center} Plots not discussed in the text of the work. \end{center}
\vskip 2ex
\centerline{
\epsfxsize=180pt
\epsfbox{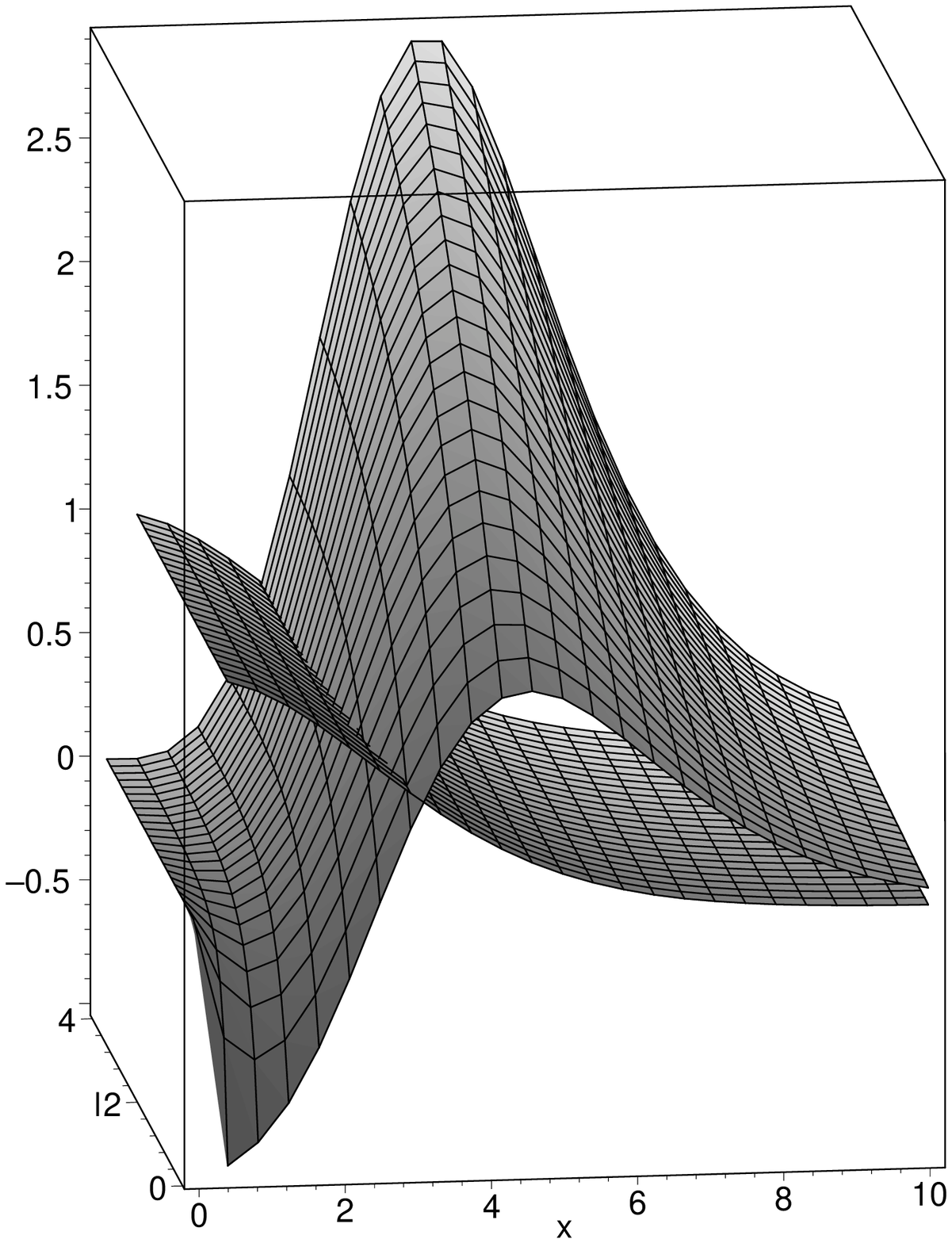}}
\vskip 4ex
\noindent
{\small{Fig. 11}.$\quad$
Heat capacity for the one-parameter Planck case compared to the standard case.}

\vskip 2ex
\centerline{
\epsfxsize=180pt
\epsfbox{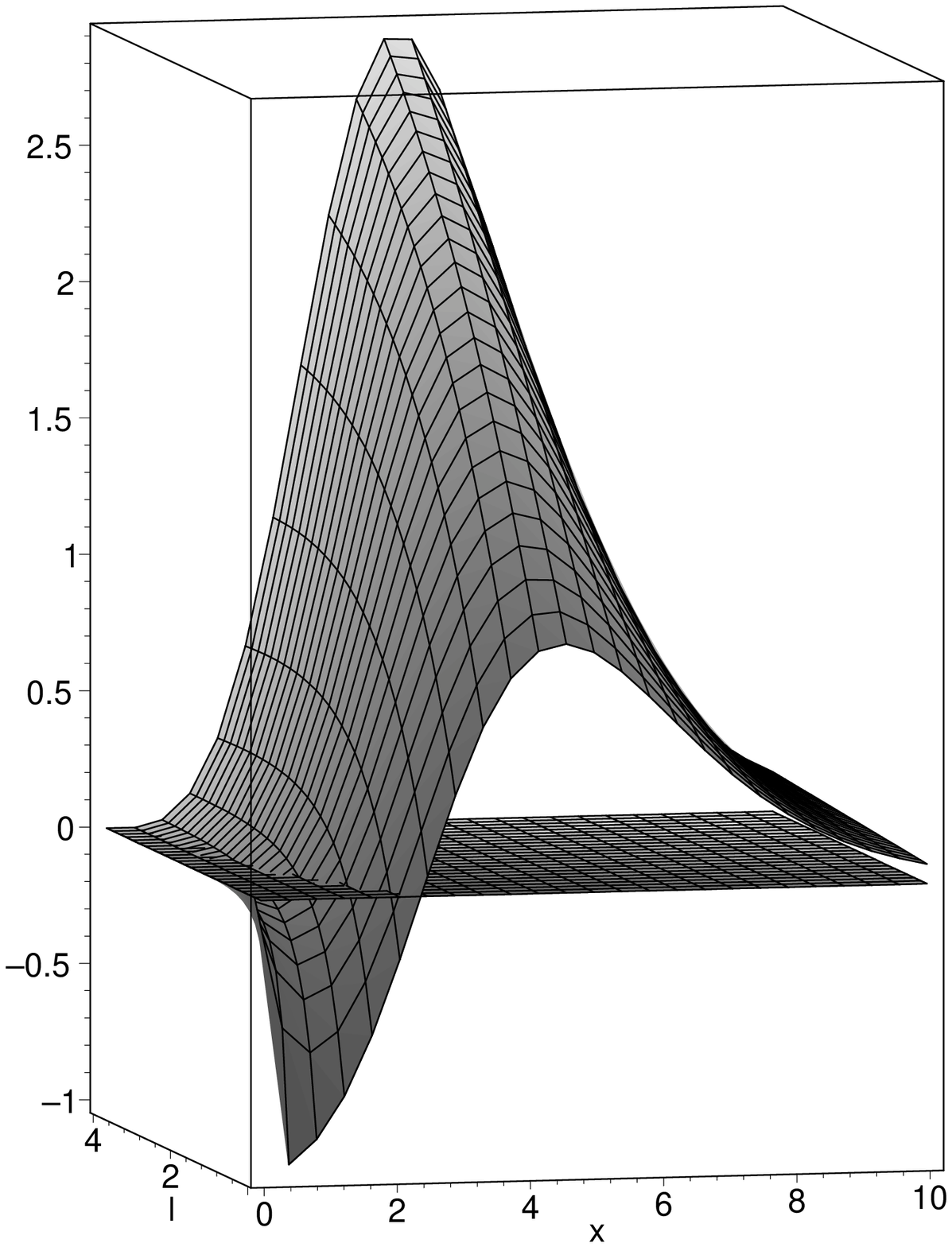}}
\vskip 4ex
\noindent
{\small{Fig. 12}.$\quad$
Heat capacity for the one-parameter vacuum case compared to the standard case.}

\vskip 2ex
\centerline{
\epsfxsize=180pt
\epsfbox{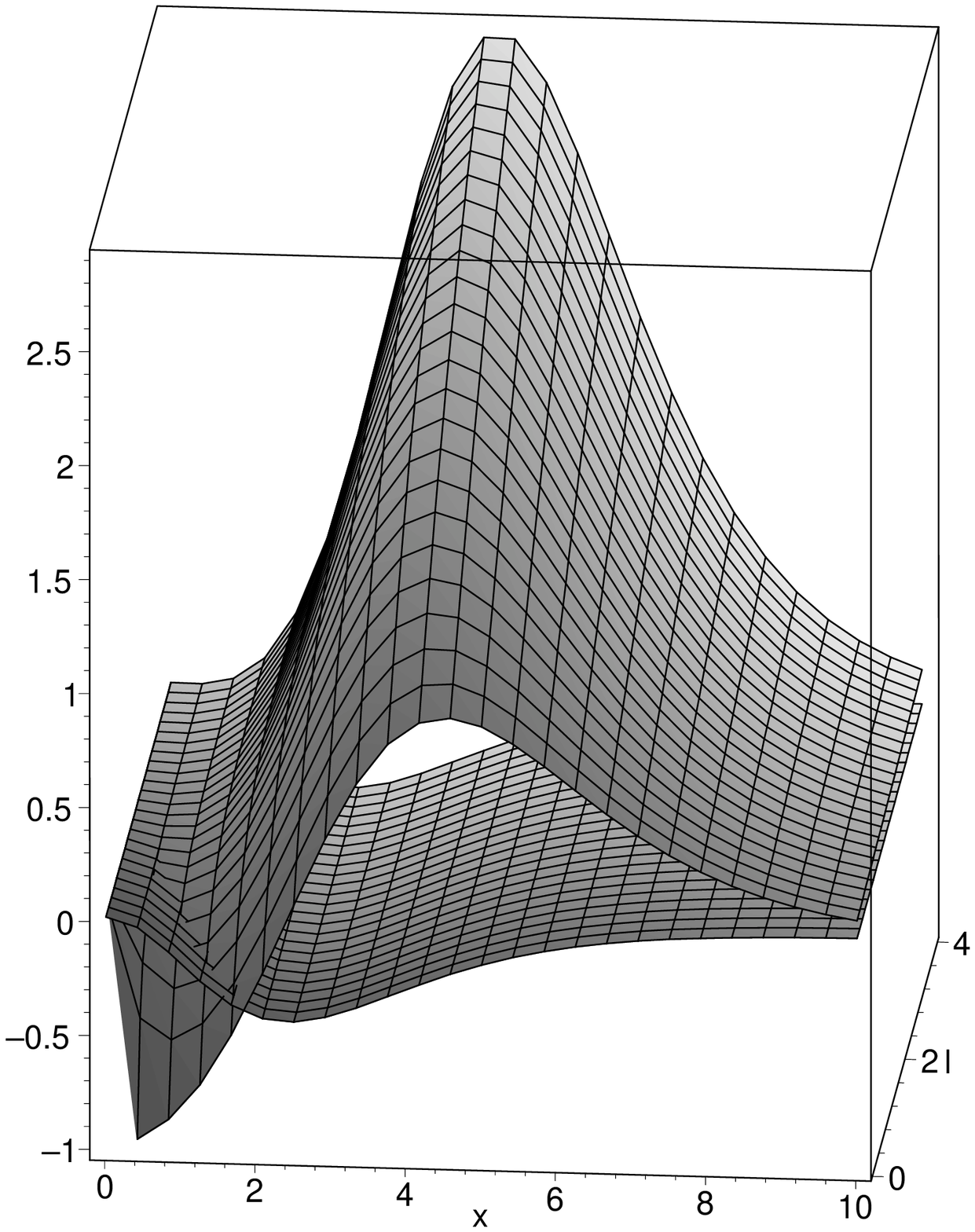}}
\vskip 4ex
\noindent
{\small{Fig. 13}.$\quad$
Heat capacity for the one-parameter Fermi-Dirac case of negative $T$ compared 
with the nonparametric case.}




\end{document}